\documentclass{article}
\usepackage[preprint]{spconf}
\usepackage{amsmath,amssymb,graphicx,hyperref}
\usepackage{caption, subfig}
\title{Selective Classifier-free Guidance for Zero-shot Text-to-speech}
\name{John Zheng, Farhad Maleki}
\address{University of Calgary, Department of Computer Science, Canada}
\begin{document}
\ninept

\setlength{\abovecaptionskip}{1ex}
\setlength{\belowcaptionskip}{1ex}
\setlength{\floatsep}{1ex}
\setlength{\textfloatsep}{1ex}

\maketitle
\begin{abstract}
In zero-shot text-to-speech, achieving a balance between fidelity to the target speaker and adherence to text content remains a challenge. While classifier-free guidance (CFG) strategies have shown promising results in image generation, their application to speech synthesis are underexplored. Separating the conditions used for CFG enables trade-offs between different desired characteristics in speech synthesis. In this paper, we evaluate the adaptability of CFG strategies originally developed for image generation to speech synthesis and extend separated-condition CFG approaches for this domain. Our results show that CFG strategies effective in image generation generally fail to improve speech synthesis. We also find that we can improve speaker similarity while limiting degradation of text adherence by applying standard CFG during early timesteps and switching to selective CFG only in later timesteps. Surprisingly, we observe that the effectiveness of a selective CFG strategy is highly text-representation dependent, as differences between the two languages of English and Mandarin can lead to different results even with the same model. 
\end{abstract}

\setlength{\abovedisplayskip}{3pt} 
\setlength{\belowdisplayskip}{3pt} 
\setlength{\belowdisplayshortskip}{3pt}
\section{Introduction}
\label{sec:intro}
\textbf{Classifier-free guidance (CFG)}~\cite{ho2022classifierfreediffusionguidance} is a key component of iteratively denoising generative models such as diffusion or flow matching. 
Flow matching~\cite{lipman2022flow}---originally tested in image generation---has been used successfully in zero-shot text-to-speech (TTS) beginning with Voicebox~\cite{le2024voicebox}. Other state-of-the-art (SOTA) zero-shot TTS models continue to utilize flow matching such as F5-TTS~\cite{chen2025f5ttsfairytalerfakesfluent}, CosyVoice 3~\cite{du2025cosyvoice3inthewildspeech}, MegaTTS 3~\cite{jiang2025megatts3sparsealignment}, and Minimax-Speech ~\cite{zhang2025minimaxspeechintrinsiczeroshottexttospeech}.

Techniques inspired by CFG are also used in both diffusion and flow matching{\textendash}based image generation models, such as weight schedules~\cite{wang2024analysisclassifierfreeguidanceweight}, different types of negative prompting~\cite{armandpour2023reimaginenegativepromptalgorithm,ban2024understandingimpactnegativeprompts,koulischer2025dynamicnegativeguidancediffusion}, modifications to the CFG algorithm~\cite{chung2024cfgmanifoldconstrainedclassifierfree,fan2025cfgzeroimprovedclassifierfreeguidance,karras2024guidingdiffusionmodelbad}, or converting CFG to training-time target modification~\cite{tang2025diffusionmodelswithoutclassifierfreeguidance}. The last technique of training-time target modification has been already been applied to speech synthesis~\cite{liang2025ttswithoutclassifierfreeguidance}, but the other methods have not. We reduce this gap in the literature by evaluating several weight schedules~\cite{wang2024analysisclassifierfreeguidanceweight} and the methods proposed in CFG-Zero* on a zero-shot TTS model.

One CFG strategy used for speech synthesis in VoiceLDM~\cite{lee2023voiceldmtexttospeechenvironmentalcontext}, DualSpeech~\cite{yang2024dualspeechenhancingspeakerfidelitytextintelligibility}, and MegaTTS 3\cite{jiang2025megatts3sparsealignment} is that of selectively emphasizing different input conditions by using separate CFG weights. This allows a user to trade-off between different output characteristics. However, these techniques require additional model evaluations during inference, and degradation in the non-emphasized condition remains undesirable. In this study, we propose applying regular CFG during early timesteps and switching to separated CFG to emphasize speaker conditioning in later timesteps. This approach does not require additional model evaluations compared to regular CFG. 

However, we find that model and language differences affect which CFG strategies are effective. Our proposed strategy is effective for F5-TTS~\cite{chen2025f5ttsfairytalerfakesfluent} in English, but for Mandarin with the same model, selective CFG does not improve speaker similarity. With CosyVoice 2~\cite{du2024cosyvoice2scalablestreaming}, we find that selective CFG simply improves speaker similarity without any degradation in text adherence, so our proposed strategy is unnecessary.

\section{Background}
\label{sec:background}
\subsection{Zero-shot Text-to-Speech}
Zero-shot TTS is a type of voice cloning restricted to only a single sample of the speaker's voice, optionally also including a transcript of the sample. Zero-shot TTS is one of the most common forms of controllable TTS~\cite{xie2025controllablespeechsynthesisera} and can be further combined with other functionalities such as emotion control or pitch control.

Two standard objective metrics for evaluating zero-shot TTS systems are similarity score (SIM) and word error rate (WER). SIM is measured by using a speaker verification system such as WavLM Large~\cite{chen2022wavlm} to extract speaker embeddings from both the reference audio and the generated audio, with the cosine similarity between the two embeddings reported as SIM. WER is calculated by comparing the input text with the results of an automatic speech recognition (ASR) model on the generated audio and calculating the percentage of words that are added, removed, or substituted. SIM and WER are useful metrics for zero-shot TTS systems as they individually measure adherence to the reference audio and the input text, respectively. 
%% However, SIM and WER are limited by the capabilities of the models used to calculate the scores, so a ground-truth SIM and WER (computed using a second utterance from the same speaker) are usually included in reported results.
\subsection{Flow Matching}
Flow matching~\cite{lipman2022flow} is a method of efficiently training continuous normalizing flows. While flow matching was proposed as a general framework for modeling smooth transformations between distributions, optimal transport (OT) flow matching is proposed as the transformation with the straightest path through the data space. This allows OT to maintain fidelity while requiring less inference steps than other iteratively denoising algorithms such as diffusion. When applied in generative models, flow matching primarily refers to the use of OT flow matching. 

The training and inference algorithms for OT flow matching are as follows. Let $x_1$ be the original audio signal and $x_0$ be a sample of Gaussian noise. During training, a random timestep $t \in [0,1]$ is sampled, and the model receives as input $x_t = x_0 + t\times(x_1 - x_0)$ and embedding for $t$. The prediction target is $x_1 - x_0$ with $L_2$ loss, which is interpreted as the derivative with respect to $t$ of a linear function between $(x_0, t=0)$ and $(x_1, t=1)$. The inference algorithm is an ordinary differential equation (ODE) from the initial value of pure noise $x_0$ at $t = 0$ to the final value of predicted speech at $\hat{x}_1$ at $t = 1$ by integrating from $t = 0$ to $1$ with an ODE solver, using the model prediction as the derivative. 

% Continuous normalizing flows model the smooth transformation of one probability distribution to another probability distribution in the data space $\mathbb{R}^d$, such as from Gaussian noise to the distribution of audio spectrograms. This smooth transformation occurs through timesteps from $t = 0$ to $t = 1$. A corresponding vector field in $\mathbb{R}^d$ parameterized by $t$ represents how the distribution changes at time $t$. Flow matching suggests to train the model by minimizing the prediction losses of individual points in the vector field instead of modeling the entire vector field. This leads to a ``simulation-free'' method of training flow matching models, which can include diffusion models under the same framework. 

%%%%%% PUT THE ALGORITHM HERE %%%%%%%%%

\subsection{Classifier-free Guidance}
Classifier-free guidance (CFG)~\cite{ho2022classifierfreediffusionguidance}, originally introduced in diffusion models, involve amplifying the difference between a conditioned and unconditioned prediction to increase the effect of the conditioning. Given $\lambda$ as the constant CFG weight, it can be implemented as replacing the conditioned model prediction $\epsilon(x_t, c)$ with the following:
\begin{equation*}
\hat\epsilon(x_t, c) = \epsilon(x_t, c) + \lambda(\epsilon(x_t, c) - \epsilon(x_t))
\end{equation*}
where $\epsilon(x_t)$ is the unconditioned model prediction. The same concept can be applied to flow matching, where the predicted derivative $\epsilon(x_t, c)$ is replaced with $\epsilon(x_t, c) + \lambda(\epsilon(x_t, c) - \epsilon(x_t))$. CFG was originally formalized as creating an implicit classifier based on prior work with classifier guidance~\cite{dhariwal2021diffusionmodelsbeatgans}, and explanation can be found in the original~\cite{ho2022classifierfreediffusionguidance}.
%%%%%%% INSERT DIAGRAMS %%%%%%%%%%%%%%%%

\subsection{CFG for Image Generation}
Negative prompting is an extension of concepts introduced by CFG, originally used in image generation models such as Stable Diffusion and explored in other academic works~\cite{armandpour2023reimaginenegativepromptalgorithm,ban2024understandingimpactnegativeprompts}. Negative prompting is used to prevent the generation of certain features in image generation by replacing the unconditioned prediction in CFG with a prediction conditioned on the unwanted feature. With $c^-$ as the unwanted condition, the modified prediction becomes: 
\begin{equation*}
\hat\epsilon(x_t, c, c^-) = \epsilon(x_t, c) + \lambda(\epsilon(x_t, c) - \epsilon(x_t, c^-)).
\end{equation*}
It was found that negative prompts have different effects on across timesteps~\cite{ban2024understandingimpactnegativeprompts}, with negative prompts restricted to early timesteps potentially \emph{introducing} the unwanted feature to an image that otherwise did not contain it, successfully removing features if used in middle timesteps, and largely having no effect if used only in later timesteps.

Perp-Neg~\cite{armandpour2023reimaginenegativepromptalgorithm} and CFG-Zero*~\cite{fan2025cfgzeroimprovedclassifierfreeguidance} both calculate the perpendicular component between the two prediction vectors instead of just the difference between them. Perp-Neg uses the component of the negative prediction perpendicular to the positive prediction, while CFG-Zero* uses the component of the positive prediction perpendicular to the unconditioned prediction. (CFG-Zero* uses a different formalization, but the resulting algorithm is identical to what we describe.) CFG-Zero* also suggests the zero-init strategy, where ignoring the first few update steps during flow matching (i. e. beginning flow matching from $t > 0$ while still using pure noise as the starting point) may improve generation results, especially for underfitted models. 

Wang et al.~\cite{wang2024analysisclassifierfreeguidanceweight} analyzed different CFG weight schedules for image generation and finds that linearly decreasing weight schedules, optionally clamped to not decrease below a lower bound at later timesteps, can have beneficial results. 

\subsection{Separated-condition CFG for Speech Synthesis}

VoiceLDM~\cite{lee2023voiceldmtexttospeechenvironmentalcontext} uses two conditions for environmental speech synthesis, the script for the generated speech and a description of the environmental context. It uses two separate CFG weights for the text script condition $c_{text}$ and the description condition $c_{desc}$, with the guided prediction as follows:
\begin{align*}
\hat\epsilon(x_t, c_{text}, c_{desc}) &= \epsilon(x_t, c_{text}, c_{desc})\\
    &+ \lambda_{text}(\epsilon(x_t, c_{text}) - \epsilon(x_t))\\
    &+ \lambda_{desc}(\epsilon(x_t, c_{desc}) - \epsilon(x_t))
\end{align*}
DualSpeech~\cite{yang2024dualspeechenhancingspeakerfidelitytextintelligibility} uses a similar formulation, replacing the environmental description conditioning with speaker conditioning $c_{spk}$. 
\begin{align*}
\hat\epsilon(x_t, c_{text}, c_{spk}) &= \epsilon(x_t, c_{text}, c_{spk})\\
    &+ \lambda_{text}(\epsilon(x_t, c_{text}) - \epsilon(x_t))\\
    &+ \lambda_{spk}(\epsilon(x_t, c_{spk}) - \epsilon(x_t))
\end{align*}
Notably, Mega-TTS 3~\cite{jiang2025megatts3sparsealignment} changes the separated-condition CFG from DualSpeech by adding text conditioning to both conditioned and unconditioned predictions for emphasizing speaker conditioning. It uses the following formulation:
\begin{align*}
\hat\epsilon(x_t, c_{text}, c_{spk}) &= \underline{\epsilon(x_t)}\\
    &+ \lambda_{text}(\epsilon(x_t, c_{text}) - \epsilon(x_t))\\
    &+ \lambda_{spk}(\underline{\epsilon(x_t, c_{text}, c_{spk}) - \epsilon(x_t, c_{text})})
\end{align*}
with changes from DualSpeech underlined.

The authors of Mega-TTS 3 find that as $\lambda_{text}$ increases, it starts with poor text adherence at $\lambda_{text}=1$, shifts towards accented pronunciation around $\lambda_{text}=1.5$ to $\lambda_{text}=2.5$, and then finally towards standard pronunciation at $\lambda_{text}=4$ and beyond ~\cite{jiang2025megatts3sparsealignment}. 

\section{Text-to-speech Systems}
\label{sec:ttssystems}
We investigate two strong (SOTA or near-SOTA) open weight models, F5-TTS~\cite{chen2025f5ttsfairytalerfakesfluent} and CosyVoice 2~\cite{du2024cosyvoice2scalablestreaming}. We selected these models primarily based on availability, as many other state-of-the-art zero-shot TTS models are not publicly available (most notably Seed-TTS~\cite{anastassiou2024seedttsfamilyhighqualityversatile}). In addition, these models represent the two most popular flow-matching TTS paradigms: purely non-autoregressive flow matching and flow matching on autoregressively generated speech tokens. 

F5-TTS is based on E2-TTS~\cite{eskimez2024e2ttsembarrassinglyeasy}, primarily trained with the open YouTube-based dataset Emilia~\cite{he2024emiliaextensivemultilingualdiverse} and utilizing an updated architecture and sampling strategy compared to E2-TTS. The model utilizes input audio and input text as conditioning, where the input audio is an audio clip of the voice to be cloned, and the input text is the transcript of the input audio concatenated with the desired speech text. The best-performing released checkpoint of this model has 336 million parameters. The version of the model included in the original paper has 0.66 SIM score and 0.024 WER on LibriSpeech, but later updates to the model improve it to 0.676 SIM and 0.020 WER. We consider the best iteration of this model as open-weight because the exact training or fine-tuning methodology of the updated model checkpoint is unpublished. The model uses a text embedder and a Diffusion Transformer backbone with 4.3 and 332 million parameters, respectively.

CosyVoice 2~\cite{du2024cosyvoice2scalablestreaming} is the second generation of CosyVoice models. There is a newer CosyVoice 3~\cite{du2025cosyvoice3inthewildspeech} by the same team, but as CosyVoice 3 is not public, we use the older CosyVoice 2 for our experiments. CosyVoice 2 utilizes a pre-trained LLM, Qwen2.5-0.5B, that has been fine-tuned to autoregressively generate the intermediate representations of an ASR model based on the input text. The intermediate representations of the ASR model, referred to as semantic tokens, capture the text information. The reference audio and transcript are processed into a speaker embedding. The semantic tokens and speaker embedding are used as input to a 71 million--parameter flow matching model. 

Both models use a cosine scheduler, where timesteps are not linearly distributed. For $n$ total inference timesteps, the $i$th timestep is determined by the following equation:
\begin{equation*}
    t_i = 1 - \cos \left(\frac{\pi}{2n} \times (i-1) \right) 
    \label{swaysampling}
\end{equation*}
This biases inference towards smaller updates at the start, when noise levels are high. F5-TTS defaults to 32 timesteps while CosyVoice 2 defaults to 10 timesteps. Both models perform flow matching with mel-spectrograms, and the generated mel-spectrogram is then decoded by a vocoder into audio. 

\section{Methodology}
\label{sec:methodology}
%Experiments are performed with F5-TTS~\cite{chen2025f5ttsfairytalerfakesfluent} or CosyVoice 2~\cite{du2024cosyvoice2scalablestreaming}. 
%SIM and WER scores are evaluated on the cross-sentence task.

We propose three selective CFG strategies. In the first, the input\_text condition replaces the unconditioned prediction with a prediction partially conditioned on the input text, as shown below:
\begin{align*}
\hat\epsilon(x_t, c_{text}, c_{spk}) &= \epsilon(x_t, c_{text}, c_{spk}) \\
&+ \lambda(\epsilon(x_t, c_{text}, c_{spk}) - \epsilon(x_t, c_{text})).
\end{align*}
It is also equivalent to setting $\lambda_{text} = 1$ in the separated-condition CFG used by MegaTTS 3. 

The second condition we propose is def\_text, which uses standard CFG for early timesteps below a certain threshold $t_{threshold}$ and uses the same strategy as input\_text for all timesteps above $t_{threshold}$. This design is motivated by listening to an extrapolated generation at each timestep, where the extrapolated signal is defined as $x_t + (1-t)\epsilon(x_t, c)$. We observe that the words become audible very quickly, at around 6 timesteps with $t \approx 0.04$. We hypothesize that CFG for text adherence may not be necessary after the initial steps, since the conditioned model has already captured sufficient text information. The timestep-dependent nature of conditioning has also been reported in previous work on negative prompting~\cite{ban2024understandingimpactnegativeprompts}. We evaluated candidate values of $t_{threshold} \in \{0.02, 0.04, 0.06, 0.08\}$ on F5-TTS and found that $t_{threshold} = 0.08$ achieves a good balance between improved SIM while minimizing WER increases. $t_{threshold}$ is between the 9th and 10th timesteps for F5-TTS and the 3rd and 4th timesteps for CosyVoice 2. The input\_text and def\_text conditions are evaluated using both F5-TTS and CosyVoice 2 on LibriSpeech~\cite{panayotov2015librispeech} and the English and Mandarin subsets of Seed-TTS-eval~\cite{anastassiou2024seedttsfamilyhighqualityversatile}.

We define a third condition input\_audio which, similarly to input\_text, replaces the unconditioned prediction with a prediction partially conditioned on the input audio. This is only evaluated using F5-TTS on LibriSpeech~\cite{panayotov2015librispeech} due to its poor performance. 

For Seed-TTS-eval~\cite{anastassiou2024seedttsfamilyhighqualityversatile}, we use the provided English and Mandarin cross-sentence prompt lists.
For LibriSpeech~\cite{panayotov2015librispeech}, we use the 1127-sample prompt list provided by F5-TTS. Experiments with F5-TTS follow the original evaluation protocol of taking the average of three seeded trials~\cite{chen2025f5ttsfairytalerfakesfluent}, but differences between seeds are insignificant. For experiments with CosyVoice 2 we perform only one trial.

We also evaluate the weight schedules proposed by~\cite{wang2024analysisclassifierfreeguidanceweight}---namely the linearly increasing schedule and the clamped-minimum linearly increasing schedule---as well as the perpendicular reweighting and zero-init strategies introduced in CFG-Zero*~\cite{fan2025cfgzeroimprovedclassifierfreeguidance}, using F5-TTS on the LibriSpeech~\cite{panayotov2015librispeech} test set.

For F5-TTS, $c_{spk}$ is treated as the input audio and $c_{text}$ is treated as the transcript text concatenated with the input text. The transcript text could arguably be considered part of $c_{spk}$, but we leave this for future research. It is worth noting that F5-TTS is trained in fully conditioned, text only, and fully unconditioned modalities~\cite{chen2025f5ttsfairytalerfakesfluent}.

For CosyVoice 2, $c_{spk}$ is treated as the speaker embedding and $c_{text}$ as the output semantic tokens of the Qwen2.5-0.5B model.

\section{Results}
\label{sec:experiments}
\begin{figure}[tbph]
    \includegraphics[scale=0.5]{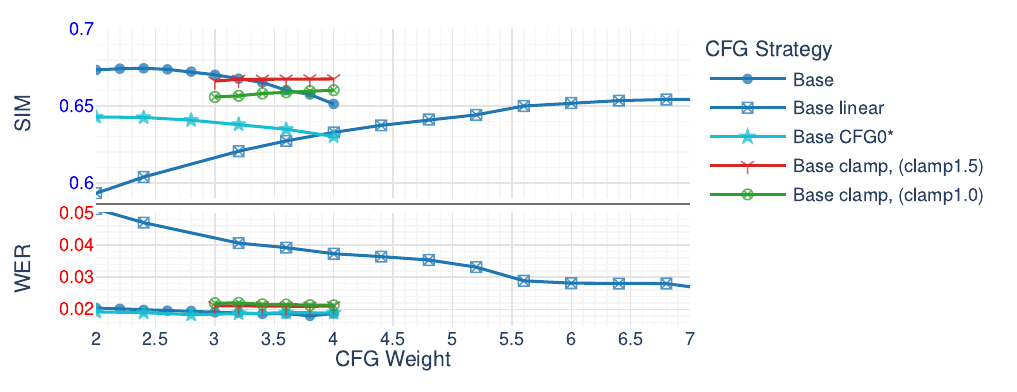}
    \caption{Results of using F5-TTS with some CFG methods which reported improved image generation quality.}
    \label{fig:repro}
\end{figure}

\begin{figure}[!tbph]
    \includegraphics[scale=0.5]{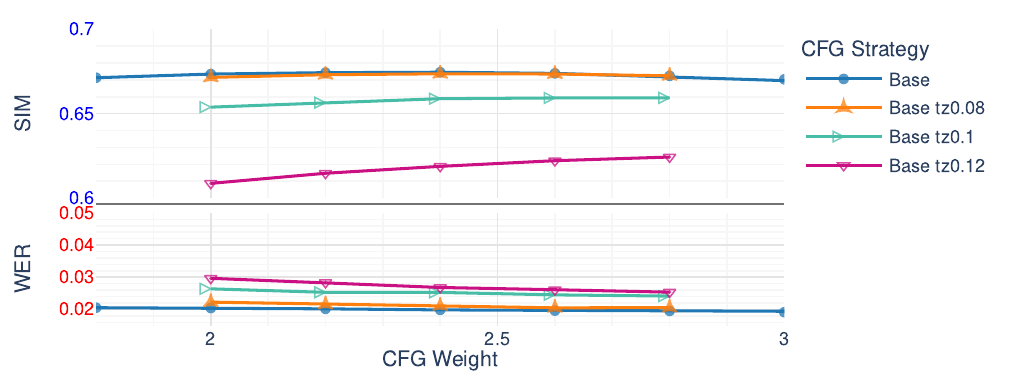}
    \caption{Evaluation of late timestep method from CFG-Zero*~\cite{fan2025cfgzeroimprovedclassifierfreeguidance}, with the value $tz$ controlling the starting timestep value. The starting timestep value is $1 - \cos (\frac{\pi}{2} \times tz)$, as the value $tz$ was used before the cosine scheduler was applied.}
    \label{fig:cfgztimestep}
\end{figure}

The weight schedules from~\cite{wang2024analysisclassifierfreeguidanceweight} and the perpendicular re-weighting from CFG-Zero*~\cite{fan2025cfgzeroimprovedclassifierfreeguidance} perform worse than the baseline, as shown in Figure \ref{fig:repro}. However, we note that high early CFG weight heavily degrade generation quality more than high CFG weight in late timesteps. Zero-init from CFG-Zero*~\cite{fan2025cfgzeroimprovedclassifierfreeguidance} is implemented by starting from a later timestep instead of skipping inference steps, and this also does not improve generation quality as shown in Figure \ref{fig:cfgztimestep}.

\begin{figure}[!tbph]
    \includegraphics[scale=0.5]{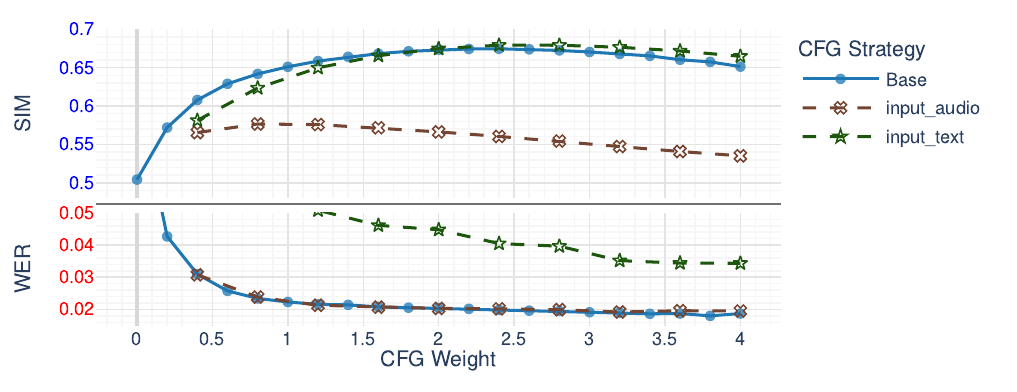}
    \caption{Adding some conditioning to the unconditioned prediction, and therefore not emphasizing those conditions with CFG.}    
    \label{base_audio_text}
\end{figure}

In Figure \ref{base_audio_text}, we find that input\_audio does not have lower WER. However, input\_text does achieve a trade-off of higher SIM at the cost of worse WER. 

\begin{figure}[tbph]
    \includegraphics[scale=0.5]{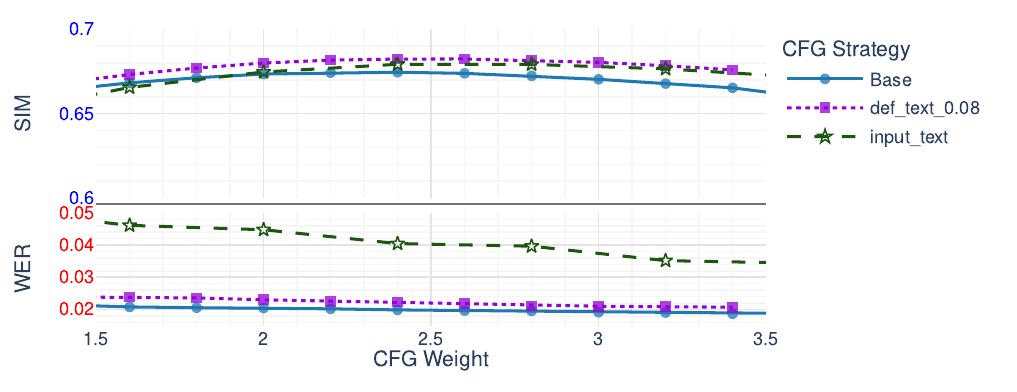}
    \caption{Comparison of baseline, def\_text, and input\_text strategies on LibriSpeech with F5-TTS.}
    \label{def_0.08}
\end{figure}

\begin{figure}[!tbph]
    \includegraphics[scale=0.5]{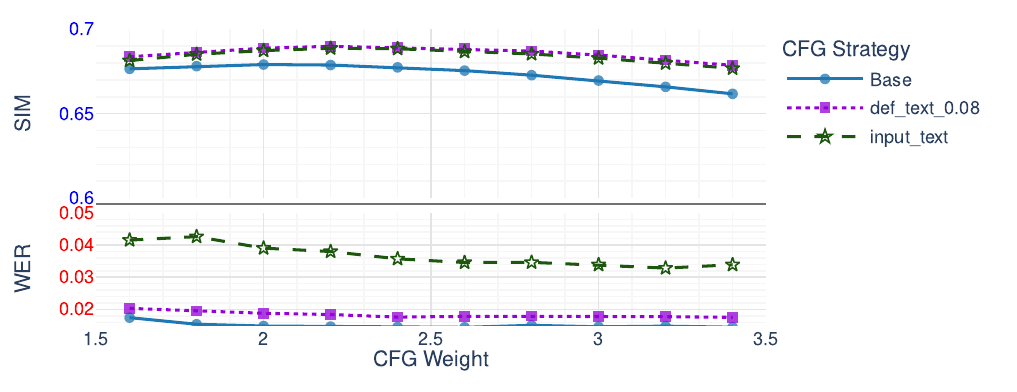}
    \caption{Results on English subset of Seed-TTS-eval with F5-TTS.}
    \label{seed_tts_en}
\end{figure}

\begin{figure}[!tbph]
    \includegraphics[scale=0.5]{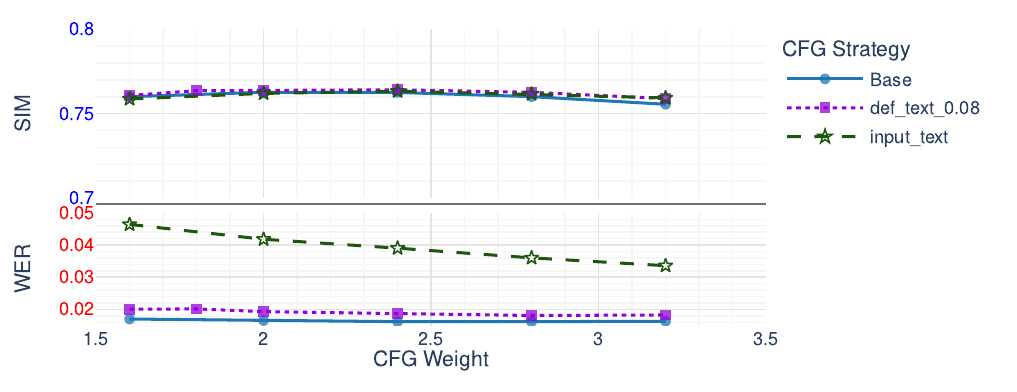}
    \caption{Results on Mandarin subset of Seed-TTS-eval with F5-TTS show no gain in SIM with input\_text or def\_text compared to baseline.}
    \label{seed_tts_zh}
\end{figure}

We find that def\_text with a threshold $t_{threshold} = 0.08$ achieves a good balance of increasing SIM with minimal impact to WER, as shown in Figure \ref{def_0.08}. This approach also works for the English subset of Seed-TTS-eval~\cite{anastassiou2024seedttsfamilyhighqualityversatile} as shown in Figure \ref{seed_tts_en}. 
However, using the exact same model, neither input\_text nor def\_text improve SIM for the Mandarin dataset and only increases WER, as shown in Figure \ref{seed_tts_zh}. 
\iffalse
Improvements to SIM score are also not consistent for every sample, as shown in Figure \ref{comparison}. We compare the performance of the baseline configuration and def\_text\_0.08 for each test case in the LibriSpeech evaluation set and find that while there is an overall improvement of 0.0070 with both at CFG strength of 2.0, individual generations have fairly high variance.

\begin{figure}[!tbph]
    \includegraphics[scale=0.5]{comparison_base_def_text.pdf}
    \caption{Histogram of the difference in SIM score between def\_text\_0.08 and the baseline.}
    \label{comparison}
\end{figure}
\fi

\begin{figure}[!tbph]
\includegraphics[scale=0.5]{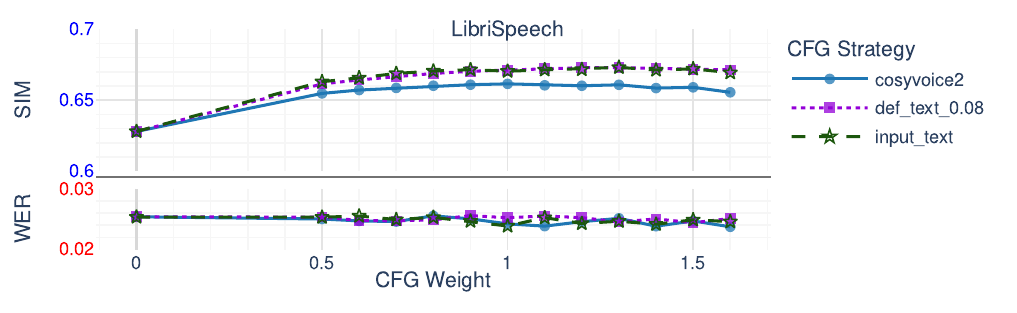}
\includegraphics[scale=0.5]{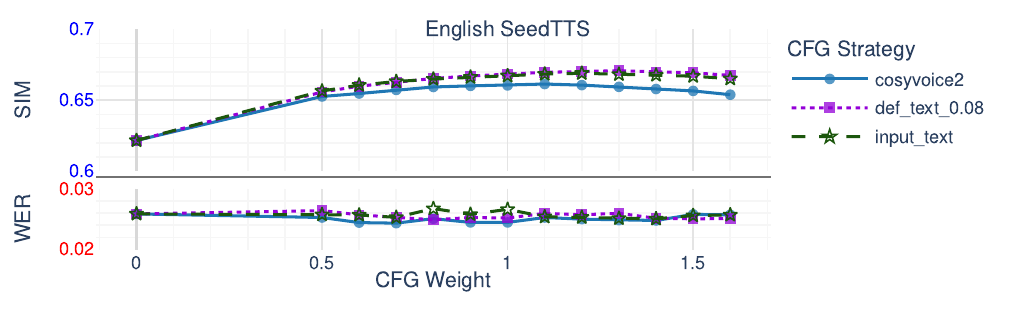}
\includegraphics[scale=0.5]{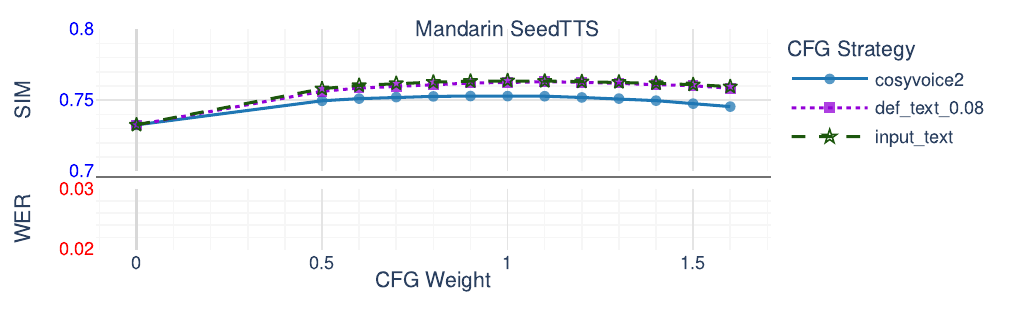}
\caption{Results for CosyVoice 2. Note that WER is low even without CFG (at CFG strength of 0).}
\label{cosyvoice}
\end{figure}

From Figure \ref{cosyvoice}, we observe that with CosyVoice 2, the input\_text condition does not degrade WER as it did with F5-TTS. We also did not observe significant language differences between English and Mandarin, unlike F5-TTS. Also we note that while CosyVoice 2 defaults to a CFG strength of 0.7, higher SIM scores appear to be achieved at CFG strength of around 1.0.

%\vfill
%\pagebreak

\section{Discussion}
\label{sec:discussion}

\begin{table}[!tbph]
    \centering
    \resizebox{\columnwidth}{!}{%
    \begin{tabular}{c|c|c|c|c|c|c}
         Model &  \multicolumn{2}{c|}{LibriSpeech}  &  \multicolumn{2}{c|}{Seed-TTS-en} &  \multicolumn{2}{c}{Seed-TTS-zh} \\
          & SIM & WER & SIM & WER & SIM & WER \\
          \hline
          F5-TTS (Base) & \textit{0.675} & \textit{0.020} & \textit{0.679} & \textit{0.018} & \textit{0.763} & \textit{0.017}\\
          F5-TTS (def\_text) & \textit{0.682} & \textit{0.022} & \textit{0.690} & \textit{0.018} & \textit{0.764} & \textit{0.019}\\
          CosyVoice 2 (Base) & \textit{0.661} & \textit{0.025} & \textit{0.660} & \textit{0.024} & \textit{0.753} & \textit{0.017}\\
          CosyVoice 2 (input\_text) & \textit{0.671} &  \textit{0.025} & \textit{0.666} & \textit{0.026} & \textit{0.763} & \textit{0.018}\\
          \hline
          Minimax-Speech~\cite{zhang2025minimaxspeechintrinsiczeroshottexttospeech} & & & 0.738 & 0.019 & 0.799 & 0.010 \\
          Seed-TTS~\cite{anastassiou2024seedttsfamilyhighqualityversatile} & & & 0.762 & 0.022 & 0.796 & 0.011 \\
          CosyVoice 3-1.5B~\cite{du2025cosyvoice3inthewildspeech} & & & 0.720 & 0.022 & 0.781 & 0.012 \\
          \hline
          
    \end{tabular}
    }
    \caption{Comparison of state-of-the-art zero-shot TTS models. Italicized results are experimentally obtained; others are reported.}
    \label{tab:sota}
\end{table} 

We observed that the tested CFG strategies developed for image generation, namely weight schedules~\cite{wang2024analysisclassifierfreeguidanceweight}, perpendicular reweighting~\cite{fan2025cfgzeroimprovedclassifierfreeguidance}, and zero-init~\cite{fan2025cfgzeroimprovedclassifierfreeguidance}, do not generalize well to zero-shot TTS with F5-TTS as shown in Figure~\ref{fig:repro}. This could be due to differences in conditioning and output modality. We suggest that future research assess the contribution of these techniques across different target modalities, such as various audio codecs~\cite{xie2025controllablespeechsynthesisera}.

Language differences have an impact on CFG strategy effectiveness for F5-TTS, but not for CosyVoice 2. This may be due to differences in text representation. CosyVoice 2 uses a 506 million--parameter LLM~\cite{du2024cosyvoice2scalablestreaming}, so generated semantic tokens allow the model to achieve strong text adherence even without CFG as seen in Figure \ref{cosyvoice}. F5-TTS relies on a much smaller 4.3 million--parameter ConvNeXt 2 module to process text, which may result in English and Chinese text acting as different conditioning modalities. However, the language differences do not arise in WER but rather through a failure to improve SIM when using either input\_text or def\_text conditions, which is not where language differences may be expected to arise. Another possible cause of the language difference for F5-TTS is training methodology or dataset differences, but as the training procedure for the best-performing checkpoint is not published, this remains speculation. 

Even though our proposed CFG strategy can improve SIM for F5-TTS and CosyVoice 2, these improvements do not completely close the gap of state-of-the-art results reported by other closed-source models as listed in Table \ref{tab:sota}. However, given that our methods can find significant improvements with only an inference-time hyperparameter sweep, they are added gains without difficulties of training large models or collecting large, high-quality datasets.

\section{Conclusion}
\label{sec:conclusion}
Here, we confirm prior results~\cite{lee2023voiceldmtexttospeechenvironmentalcontext,yang2024dualspeechenhancingspeakerfidelitytextintelligibility,jiang2025megatts3sparsealignment} that a trade-off between speaker similarity and text adherence can be achieved by separately emphasizing the speaker conditioning with CFG. We show that by starting with regular CFG and switching to selective CFG after the several timesteps, as inspired by prior investigations into negative prompting~\cite{ban2024understandingimpactnegativeprompts}, may minimize WER increases while still improving SIM. However, we demonstrate that the efficacy of separated-condition CFG depends on both the language and the model being used.

% \vfill\pagebreak
% \section{Compliance with Ethical Standards}
% This research was supported by the Natural Sciences and Engineering Research Council of Canada (NSERC), Discovery Grant RGPIN-2024-04966. The authors have no other relevant financial interests to disclose. This study was conducted using code and model checkpoints from F5-TTS~\cite{chen2025f5ttsfairytalerfakesfluent}, code and model checkpoints from CosyVoice 2~\cite{du2024cosyvoice2scalablestreaming}, Seed-TTS-eval~\cite{anastassiou2024seedttsfamilyhighqualityversatile}, and LibriSpeech~\cite{panayotov2015librispeech} under public domain, share-alike, or non-commercial licenses.

\bibliographystyle{IEEEbib}
\bibliography{refs}

\end{document}